\documentstyle[aps,preprint]{revtex}
\newcommand{\be}{\begin{equation}}
\newcommand{\ee}{\end{equation}}
\newcommand{\beq}{\begin{eqnarray}}
\newcommand{\eeq}{\end{eqnarray}}
\newcommand{\mbf}{\mathbf}

\newcommand{\etab}{\mbox{\boldmath$\eta$}}
\begin{document}
\tightenlines
\title{Generalized Turing Patterns and Their Selective Realization in
Spatiotemporal Systems}
\author{Govindan Rangarajan$^{1}$, Yonghong Chen$^2$
and Mingzhou Ding$^{2}$}

\address{$^1$Department of Mathematics and Center for Theoretical
Studies, Indian Institute of Science, Bangalore 560 012, India}
\address{$^2$Center for Complex Systems and Brain Sciences,
Florida Atlantic University, Boca Raton, FL 33431, USA}

\maketitle

\begin{abstract}
We consider the pattern formation problem in coupled identical
systems after the global synchronized state becomes unstable.
Based on analytical results relating the coupling strengths and
the instability of each spatial mode (pattern) we show that these
spatial patterns can be selectively realized by varying the
coupling strengths along different paths in the parameter space.
Furthermore, we discuss the important role of the synchronized
state (fixed point versus chaotic attractor) in modulating the
temporal dynamics of the spatial patterns.
\end{abstract}

{\it Introduction.} In a reaction-diffusion system, at the Turing
bifurcation point the global equilibrium state becomes unstable
and an inhomogeneous spatial pattern known as a Turing pattern
\cite{turing} emerges. Following Turing's classic work pattern
formation in reaction-diffusion systems has become a major topic
of investigation both theoretically and experimentally
\cite{murray,kaneko}.

The idea in Turing's original formulation is not limited to just
reaction-diffusion systems with a single diffusion constant.
Networks where the diffusion is characterized by a more complex
function and networks where the coupling is not diffusive can be
studied in a similar fashion. This makes the idea of pattern
formation applicable to diverse areas of science and engineering
\cite{ding,kocarev,klev,li1,wu,cuomo,pecora,rd}. Furthermore, the
globally synchronized state prior to the formation of spatial
patterns need not be a fixed point as in Turing's original case.
One can consider synchronized limit cycles and chaotic attractors
as the bifurcating state. When the synchronized state becomes
unstable under parameter variation, the dynamics may exit the
synchronization manifold along certain eigen-mode, giving rise to
a spatial pattern. We refer to these patterns as Generalized
Turing Patterns (GTPs).

GTPs have been the subject of several recent publications
\cite{pecora}. These papers mainly rely on numerical techniques to
obtain the threshold of instabilities. In this letter we focus on
the analytical treatment of the pattern formation and pattern
selection. To do so we restrict ourselves to networks in which
each node is coupled to its $P$ nearest neighbors. Under the
assumption of periodic boundary condition and translational
invariance the eigen-modes are the spatial Fourier modes. Our
main results are as follows. First, we derive explicit analytical
expressions defining the stability region in the parameter space
spanned by the coupling strengths. Second, we demonstrate that,
by following different paths in the parameter space, different
spatiotemporal patterns can be selectively realized. Third,
although the spatial patterns are the same for a given coupling
matrix, we show that, depending on whether we have a synchronized
fixed point or chaotic attractor, the temporal evolution of the
patterns is either constant in time or modulated in an on-off
intermittent fashion. We study networks on both one dimensional
(1-d) and two dimensional (2-d) lattices. Only coupled maps are
considered here. Similar results apply to coupled differential
equations.

{\it One dimensional lattices.} Consider the following 1-d
coupled map lattice where each node is coupled to its $P$ nearest
neighbors in a general diffusive manner: \be\label{map} {\mbf
x}_j(n+1)={\mbf f}({\mbf x}_j(n))+ \frac{1}{2P} \sum_{p=1}^P a_p
\left[ {\mbf f}({\mbf x}_{j+p}(n)) + {\mbf f}({\mbf x}_{j-p}(n))
-2 {\mbf f}({\mbf x}_j(n)) \right], \ \ \ j=1,2, \ldots ,L. \ee
Here the $M$-dimensional vector ${\mbf x}_j$ describes the map at
the $j$th node of the lattice, $a_p$'s denote the coupling
strengths and $n$ is the time index. We impose the periodic
boundary condition: ${\mbf x}_{j+L}(n)={\mbf x}_j(n)$. Since the
coupling strengths don't depend on the spatial position index $j$
we have translational invariance. The individual (uncoupled) map
is ${\mbf x}_j(n+1)={\mbf f}({\mbf x}_j(n))$.

Clearly the synchronized state ${\mbf x}_j={\mbf x}, \ \forall \
j$ is a solution of the above system. To study its stability we
linearize Eq. (\ref{map}) around the synchronized state $\mbf x$
to obtain: \be {\mbf z}_j(n+1) = {\mbf J}({\mbf x}(n)) {\mbf
z}_j(n)+ {\mbf J}({\mbf x}(n)) \frac{1}{2P} \sum_{p=1}^P a_p
[{\mbf z}_{j+p}(n) + {\mbf z}_{j-p}(n) -2 {\mbf z}_j(n)],
\label{mlinear} \ee where ${\mbf z}_j$ denotes the $j$th map's
deviations from ${\mbf x}$ and ${\mbf J}$ is the $M \times M$
Jacobian matrix corresponding to ${\mbf f}$. Consider the
discrete Fourier transform of ${\mbf z}_j$: \be \label{dft} {\mbf
\etab}_l (n) = \frac{1}{L} \sum_{j=1}^L \exp(-i 2 \pi j l/L)
{\mbf z}_j(n). \ee Performing the discrete Fourier transform on
the linearized equation we get \be {\mbf \etab}_l (n+1) = {\mbf
J}({\mbf x}(n)) \left[ {\mbf \etab}_l (n) + \frac{1}{2P}
\sum_{p=1}^P a_p \left( \exp(i 2 \pi p l/L) + \exp(-i 2 \pi p
l/L) -2 \right){\mbf \etab}_l (n) \right]. \ee Simplifying we
finally have \be {\mbf \etab}_l (n+1) = {\mbf J}({\mbf x}(n))
\left[ 1-\frac{2}{P} \sum_{p=1}^P a_p \sin^2 (\pi p l/L) \right]
{\mbf \etab}_l (n). \ee

We start by examining the case where the individual maps are
chaotic and ${\mbf x}$ corresponds to the synchronized chaotic
state. The synchronized state is stable if all the transverse
Lyapunov exponents are negative\cite{rd,comment}. Computing the
Lyapunov exponents $\mu_i$ of the above linearized equation we get
\be \mu_i(l) = h_i + \ln | 1- \frac{2}{P} \sum_{p=1}^P a_p \sin^2
(\pi p l/L) |, \ \ \ i=1,2, \ldots , M, \ee where $h_i$'s are the
Lyapunov exponents for the individual map ordered as $h_1 \ge h_2
\ge \cdots \ge h_M$. Note that the $l=0$ Fourier mode corresponds
to the synchronized chaotic state where the Lyapunov exponents of
the coupled system are given by the Lyapunov exponents of the
individual system. For each $l \neq 0$ value, $\mu_1(l)$ gives the
largest transverse Lyapunov exponent. Therefore, stability of the
synchronized chaotic state is ensured if $\mu_1(l) < 0$ for all $l
\neq 0$. By symmetry of the Fourier modes, we only need to
consider $\mu_1(l) < 0$ for $l = 1,2, \ldots L/2$ (if $L$ is odd,
we take $(L-1)/2$). Thus we have the stability conditions \be | 1-
\frac{2}{P} \sum_{p=1}^P a_p \sin^2 (\pi p l/L) | < \exp(-h_1), \
\ \ l = 1,2, \ldots L/2\ {\rm or}\ (L-1)/2. \ee

This set of inequalities defines a stability region in the
parameter space spanned by the $a_p$'s. By selecting a given
Fourier mode and choosing a suitable path in the parameter space
we can realize the corresponding GTP. Note that, if one considers
only the nearest neighbor diffusive coupling ($P=1$), the
parameter space is one dimensional and at most two GTPs can be
excited by varying the coupling strength even though there are
many possible modes. By enlarging the parameter space we obtain
much greater variety in terms of GTPs that can be realized. With
recent progress\cite{crd} in obtaining stability conditions for
general coupling matrices, we are currently investigating whether
even more general Turing patterns can be obtained.

As a numerical example we consider coupled logistic maps in the
chaotic regime where $f(x) = 1-a x^2$ with $a=1.9$. The maximum
Lyapunov exponent $h_1$ is 0.549. For simplicity, we restrict
ourselves to $L=5$ and $P=2$. The stability conditions for the
synchronized chaotic state are: \be -0.578 < 1-a_1 \sin^2 (\pi
l/5) - a_2 \sin^2 (2 \pi l/5) < 0.578, \ \ \ l=1,2 \ee which give
the stability region marked black in the parameter plane [Fig.
1(a)]. We call the $l=1$ mode the long wavelength (LW) pattern and
the $l=2$ mode the short wavelength (SW) pattern. The arrows
indicate the paths along which either of the two patterns can be
selected.

The main frame in Fig. 1(b) shows the temporal dynamics of the
long wavelength pattern for $a_1=0.96$ and $a_2=0.1$. Here
deviations from the synchronization manifold is approximated by
$$z_j(n)=x_j(n)-\sum_{j=1}^{L} x_j(n)/L$$
with $L=5$. To facilitate visualization, at each time step $n$, a
continuous function is splined through the six discrete nodes:
$z_1(n)$, $z_2(n)$, $\cdots$, $z_5(n)$, and $z_6(n)=z_1(n)$.
Furthermore, to overcome the distortion due to the two opposite
phases of a pattern, we monitor the deviation at a given node and
multiple the deviations at every node by a minus sign whenever the
deviation at the monitored node becomes negative.

Since the bifurcation undergone by the system at the boundary of
the stability region is the blow-out bifurcation and there is only
one attractor prior to the bifurcation, the dynamics in this case
is referred to as on-off intermittency \cite{comment,onoff}. The
temporal evolution of the deviations at a typical node is given by
the curve to the left of the main pattern frame. Its bursting
behavior is characteristic of on-off intermittency. The GTP itself
is given at the bottom of Fig. 1(b).

For $a_1=0.04$ and $a_2=1.1$ we observe the short wavelength
pattern in Fig. 1(c). The same visualization methods are used to
make this figure.

To understand how the synchronized state shapes the temporal
properties of the GTP after desynchronization we consider a case
where the synchronized state is a fixed point. Again we use the
coupled logistic maps and choose $a=0.5$. The fixed point is
$\bar{x}=0.73$ and the lyapunov exponent is $h_1=-0.31$. Still
letting $L=5, P=2$, we get the stability region defined by \be
-1.36 < 1-a_1 \sin^2 (\pi l/5) - a_2 \sin^2 (2 \pi l/5) < 1.36, \
\ \ l=1,2 \ee shown in Fig. 2(a). Following the arrows we can
realize either the long or the short wavelength patterns. Figure
2(b) gives the long wavelength pattern for $a_1=0.5$ and $a_2=2.5$
and Fig. 2(c) gives the short wavelength pattern for $a_1=2.5$ and
$a_2=0.5$. The same methods of plotting as that used for Figure 1
are used here. Comparing Figs. 2(b) and (c) with Figs. 1(b) and
(c) we see the same spatial patterns but different temporal
behaviors.

For the example in Fig. 2 the final GTPs are the new fixed points
displayed as a function of the space coordinate. Predicting the
exact location of these new fixed points requires the nonlinear
terms dropped in the linear stability analysis. Although it is
often the case that the spatial functions underlying the new fixed
points agree with the respective linear eigen-modes this is by no
means a guaranteed fact \cite{murray}. On the other hand, when the
synchronized state is chaotic, linear analysis will govern the
temporal evolution whenever the phase space trajectory comes back
to near the synchronization manifold.

{\it Two dimensional lattices.} Finally, we consider a 2-d
coupled map lattice given by \beq\label{map2d} && {\mbf
x}_{j,k}(n+1) = {\mbf f}({\mbf x}_{j,k}(n))+ \frac{1}{2P}
\sum_{p=1}^P \left\{a_p \left[ {\mbf f}({\mbf x}_{j+p,k}(n)) +
{\mbf f}({\mbf x}_{j-p,k}(n)) \right]
\right. \nonumber \\
& & \left. +b_p \left[ {\mbf f}({\mbf x}_{j,k+p}(n)) + {\mbf
f}({\mbf x}_{j,k-p}(n)) \right] -2 (a_p + b_p)  {\mbf f}({\mbf
x}_{j,k}(n)) \right\}, \ \ \ j,k=1,2, \ldots ,L. \eeq Linearizing
around the synchronized state ${\mbf x}_{j,k} = {\mbf x},\
\forall \ j,k$, we get \beq \label{mlinear2d} {\mbf z}_{j,k}(n+1)
& = & {\mbf J}({\mbf x}(n)) {\mbf z}_{j,k}(n)+ {\mbf J}({\mbf
x}(n)) \frac{1}{2P} \sum_{p=1}^P \left\{
a_p [{\mbf z}_{j+p,k}(n) + {\mbf z}_{j-p,k}(n)]+ \right. \nonumber \\
&& \left. b_p [{\mbf z}_{j,k+p}(n) + {\mbf z}_{j,k-p}(n)] -2
(a_p+b_p) {\mbf z}_{j,k}(n) \right\}, \eeq where ${\mbf z}$
denotes the deviation from the synchronized manifold. Applying the
2-d discrete Fourier transformation given by \be \label{dft2d}
{\mbf \etab}_{l,m} (n) = \frac{1}{L^2} \sum_{j=1}^L \sum_{k=1}^L
\exp(-i 2 \pi j l/L)\exp(-i 2 \pi k m/L){\mbf z}_{j,k}(n), \ee to
the linearized equation and simplifying we finally obtain \be
{\mbf \etab}_{l,m} (n+1) = {\mbf J}({\mbf x}(n)) \left[
1-\frac{2}{P} \sum_{p=1}^P \left(a_p \sin^2 (\pi p l/L)+ b_p
\sin^2 (\pi p m/L) \right) \right] {\mbf \etab}_{l,m}(n). \ee
Proceeding as before, the stability conditions for synchronized
state are given by \beq && | 1- \frac{2}{P} \sum_{p=1}^P \left(a_p
\sin^2 (\pi p l/L)+
b_p \sin^2 (\pi p m/L) \right) | < \exp(-h_1), \nonumber \\
&& \ \ \ l,m = 0,1,2, \ldots L/2\ {\rm or}\ (L-1)/2, \eeq with
$l=m=0$ excluded (as it corresponds to the synchronized manifold).

As in the 1-d lattice case, GTPs emerge when the synchronized
state loses its stability. This happens when the parameters are
varied across the stability boundary. We can select a particular
$(l,m)$ mode and have it realized by appropriately choosing the
coupling strengths $a_p$ and $b_p$ ($p=1,2, \ldots , P$). We will
again illustrate this with logistic maps coupled together on a 2-d
lattice. We set $a=1.5$. Then $h_1=0.231$. Choose $L=5$ and $P=1$
for simplicity. The stability conditions are \beq && -0.79 <
1-\frac{a_1}{2} \sin^2 (\pi l/5)-\frac{b_1}{2} \sin^2 (\pi m/5)<
0.79, \ \ \ \ l,m=0,1,2, \eeq with $l=m=0$ excluded. Figure 3(a)
depicts the stability region (black). Different edges of the
region are the instability thresholds of different modes. It is
obvious that only the (0,1), (1,0) and (2,2) modes can be observed
given two tuning parameters. By including additional coupling
strengths not considered here one can potentially make the (1,1),
(1,2) and (2,1) modes observable. Figure 3(b) shows a snapshot of
deviations from the synchronization manifold at different nodes
for $a_1=0.5$ and $b_1=0.5$. As before a continuous function is
fitted through all the nodes. Although not perfect the essential
features of a (2,2) pattern is quite clearly visible in the
snapshot. The other two patterns (0,1) and (1,0) can be similarly
realized by choosing proper parameters with respect to the
boundary of the stability region in Fig. 3(a).

We thank Jianfeng Feng for helpful comments. The work was
supported by US ONR Grant N00014-99-1-0062. GR was also supported
by the Homi Bhabha Fellowship and by DRDO and ISRO, India.

%\newpage

\newpage

\section*{Figure Captions}

\begin{description}
\item{\bf Figure 1:} Pattern selection from the synchronized
chaotic state in a 1-d map lattice ($P=2$). In (a), the region of
stable synchronization (black area) and distinct pattern selection
directions are shown. In (b), temporal evolution of the long
wavelength pattern is given with $a_1=0.96, a_2=0.1$. In (c),
temporal evolution of the short wavelength pattern with
$a_1=0.04, a_2=1.1$ is given.

\item{\bf Figure 2:} Pattern selection from the synchronized
equilibrium state in a 1-d map lattice ($P=2$). In (a), the
region of stable synchronization (black area) and distinct pattern
selection directions are shown. In (b), the long wavelength
pattern evolving as a fixed point in time is given with $a_1=0.5,
a_2=2.2$. In (c), the short wavelength pattern with $a_1=2.2,
a_2=0.5$ is shown.

\item{\bf Figure 3:} Pattern selection in a 2-d map lattice
($P=1$). In (a), the region of stable synchronization (black area)
and the directions of selecting some patterns in the parameter
space are given.  In (b), a (2,2) pattern is shown with $a_1=0.5,
b_1=0.5$.

\end{description}

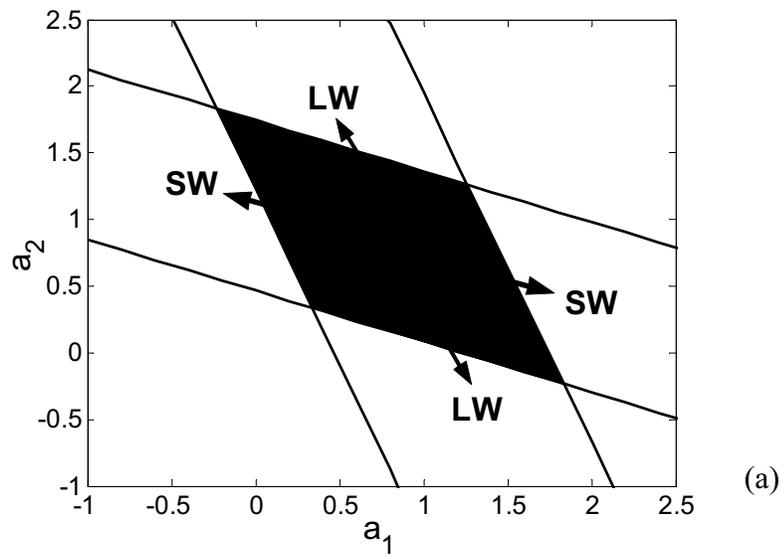
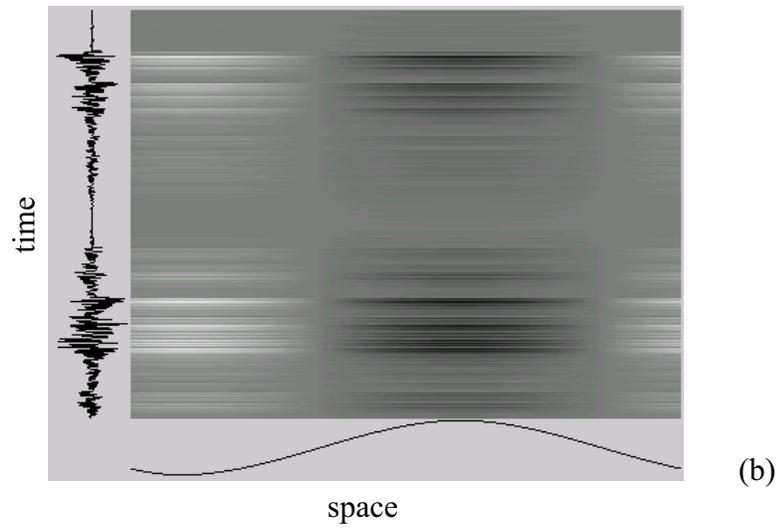
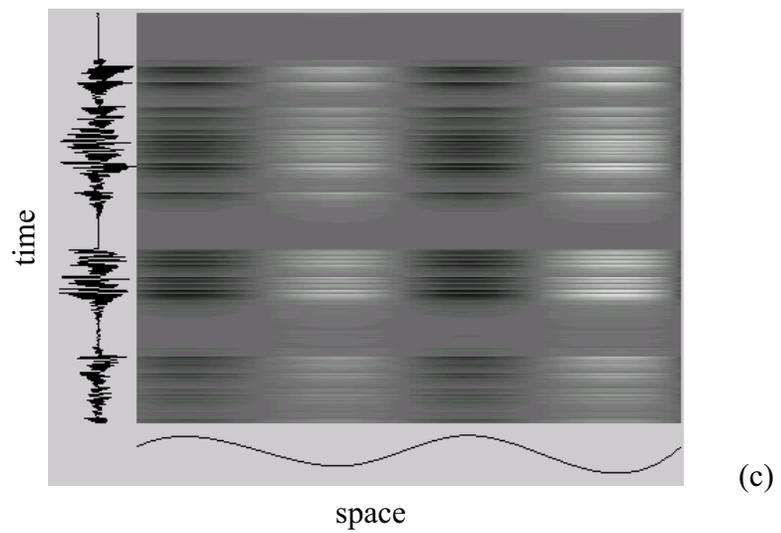
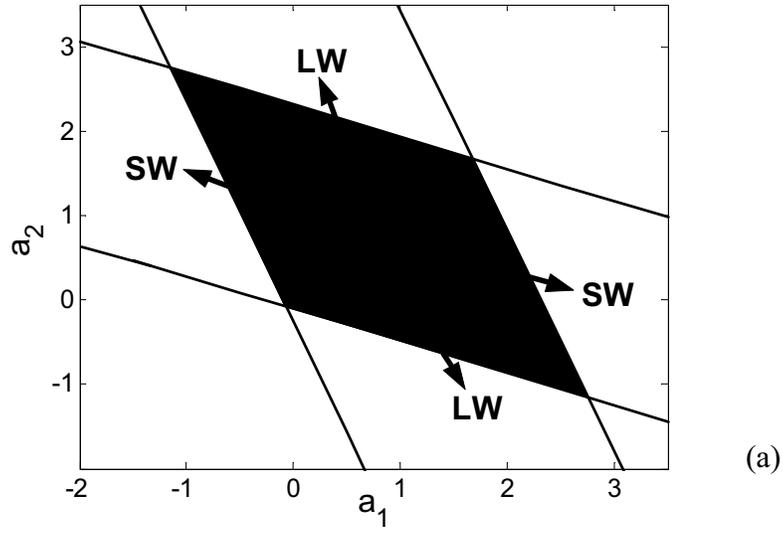
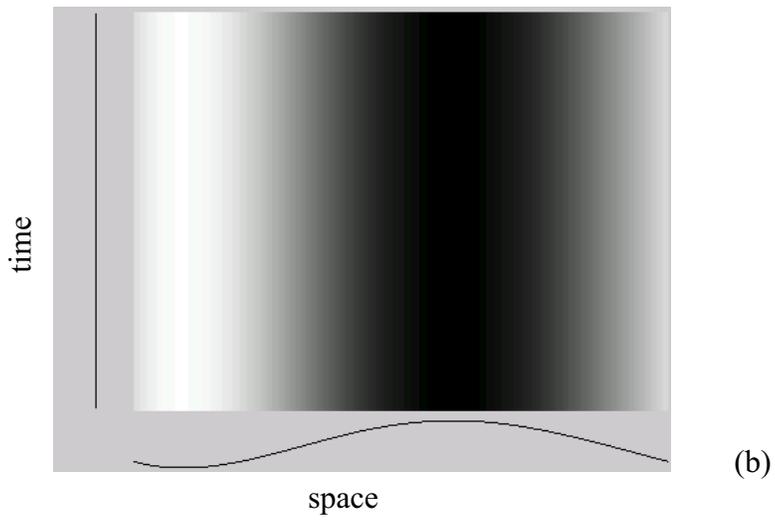
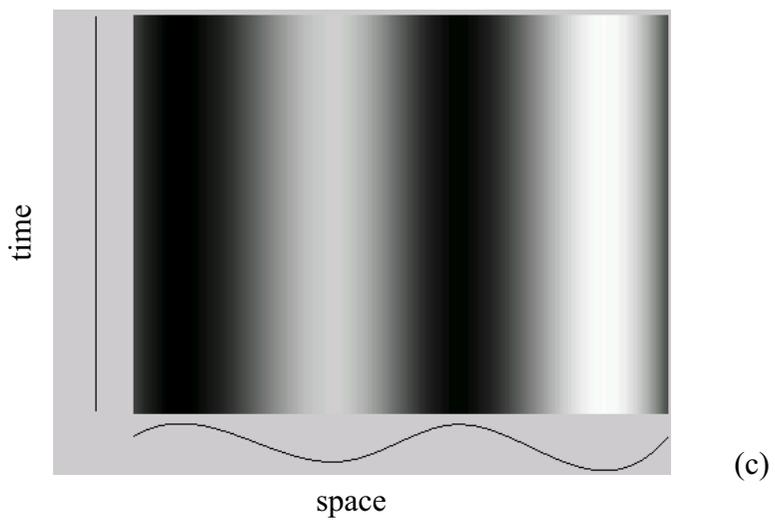
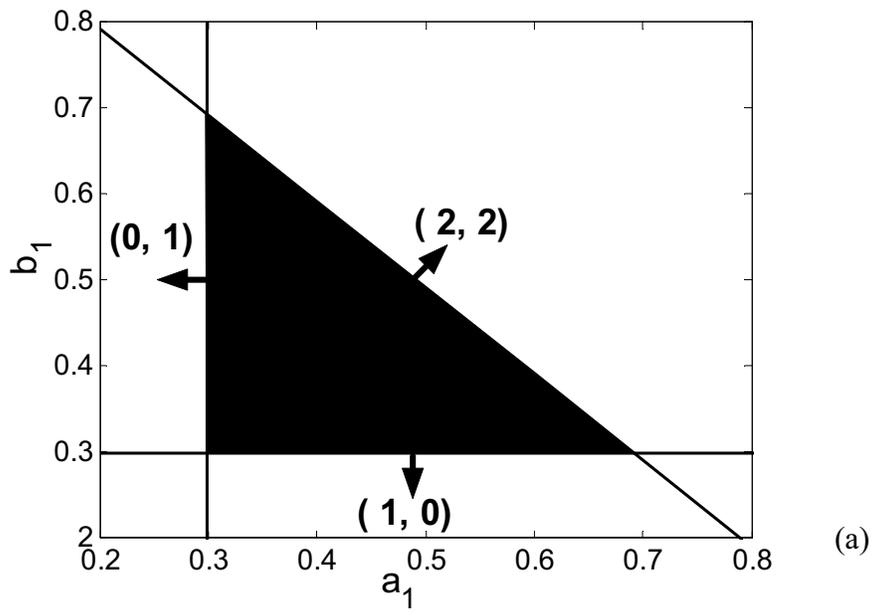
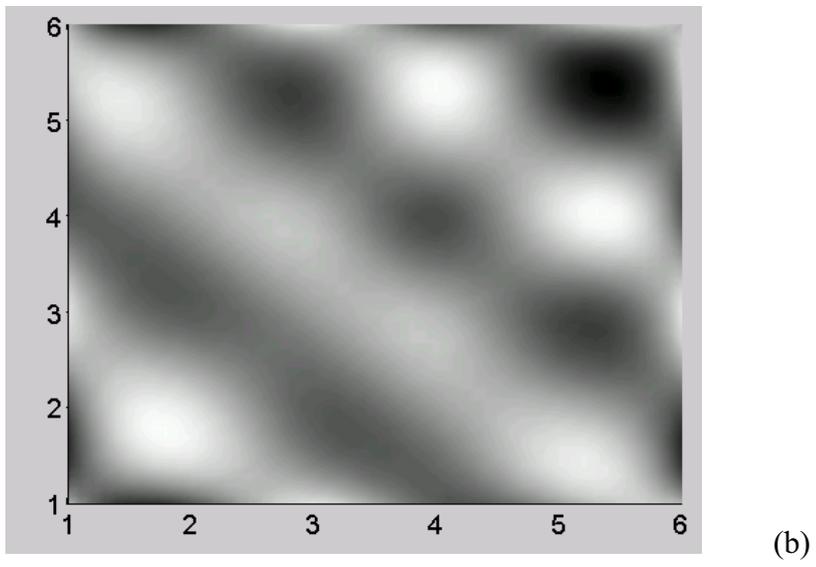
\end{document}